\documentclass[aps,pra,twocolumn,showpacs,showkeys,reprint]{revtex4-1} 
\usepackage{graphics} 
\usepackage{graphicx} 
\usepackage{amsmath} 
\usepackage{braket}
\usepackage{color}

\begin{document} 

\title{Spinor bosons in optical superlattices: a numerical study} 
\author{G. J. Cruz}
\affiliation{
Department of Physics, University at Buffalo, State University of New York, Buffalo, New York 14260, USA.
}%
\author{R. Franco} 
\author{J. Silva-Valencia} 
\email{jsilvav@unal.edu.co} 
\affiliation{Departamento de F\'{\i}sica, Universidad Nacional de Colombia, A. A. 5997 Bogot\'a, Colombia.} 

\date{\today} 

\begin{abstract} 
The ground state of spin-1 ultracold bosons trapped in a periodic one-dimensional optical superlattice is studied. The two sites of the unit cell have an energy shift between them, whose competition with the spin-dependent strength is the main focus of this paper. Charge density wave (CDW) phases appear for semi-integer and integer densities, leading to rich phase diagrams with Mott insulator, superfluid and CDW phases.
The spin-dependent interaction favors insulator phases for integer densities and disfavors CDW phases for semi-integer densities, which tend to disappear. Also, quantum phase transitions at finite values of the spin-dependent strength were observed. For integer densities, Mott insulator-superfluid-CDW insulator transitions appear for an energy shift lower (higher) than the local repulsion for the global density  $\rho=1$ ($\rho=2$).
\end{abstract} 


\maketitle 

\section{\label{sec1}Introduction} 

The emergence of the cold atom area has led to the appearance and observation of interesting physical phenomena, such as new states of matter and quantum phase transitions~\cite{IBloch-RMP08,IBloch-NP12,Stamper-RMP13}. In particular, the creation of purely optical traps with lasers unfreezes the spin degree of freedom of alkaline atoms, allowing one to observe a Bose-Einstein condensate for each hyperfine state~\cite{StamperKurn-PRL98}, spin domains~\cite{Stenger-Nature98}, coherent spin dynamics~\cite{Chang-NP05}, Larmor precession~\cite{Higbie-PRL05}, and spontaneous symmetry breaking ~\cite{Sadler-Nature06}, among other phenomena. The above observations have turned spinor bosons into a subject of great interest, which can be described by the $S$-1 Bose-Hubbard model, which considers the kinetic energy, a local two-body repulsion, and a local spin-dependent interaction~\cite{Imambekov-PRA03}. For antiferromagnetic coupling, an even-odd asymmetry in the Mott lobes appears; i.e. for an even global 
density in the system the Mott lobes grow as the spin parameter increases, while odd global density decreases. Also, it has been shown that the  odd lobes exhibit a dimerized order, while the even lobes exhibit competition between a nematic phase and a spin singlet one~\cite{Yip-PRL03,Imambekov-PRL04,Tsuchiya-PRA04,Rizzi-PRL05,Kimura-PRL05,Bergkvist-PRA06,Apaja-PRA06,Pai-PRB08,Toga-JPSJ12, Kimura-PRA13,Natu-PRB15,Li-PRA16,Hincapie-JPCS16,Hincapie-PRA16,Hincapie-PLA18}.\par 
The high degree of control of optical lattices allows confining atoms in lattices with diverse spatial configurations; among them is the superlattice, whose arrangement is characterized by a periodic potential~\cite{Roati-N08,Atala-NP13,Windpassinger-RPP13}. Also, a superlattice potential with an energy offset between two sites, A and B, has been generated in square and honeycomb lattices, where bosons and fermions were loaded, respectively~\cite{DiLiberto-NC14,Messer-PRL15}. Different ground states were observed, and transitions from superfluid to insulator can be manipulated.\par 
Spinless bosons in non-homogeneous lattices have been considered by several authors, using diverse analytical and numerical methods~\cite{Batrouni-PRL02,Buonsante-PRA04,Rousseau-PRB06,Roux-PRA08,Cazalilla-RMP11,Dhar-PRA11a,Dhar-PRA11b,Singh-PRA12,Li-PRB15,Cruz-JPCS16,Cruz-EPJB16}. From a general point of view, these studies found that the ground state can be diverse, as a function of the parameters; for instance, charge density wave (CDW), Mott insulator, superfluid or ``Bose glass" phases have been reported. In particular, for bosons confined in $AB_{n-1}$ chains, i.e. a lattice  that consists of repeating a unit cell with $n$ sites where between the $A$ and $B$ sites there is an energy offset $\lambda$, insulator phases for densities $\rho=\alpha/n$ were found, with $\alpha$ being an integer, and these insulator phases are separated by superfluid regions. For any $n$ value, it has been reported that for integer densities $\rho$, the system exhibits $\rho +1$ insulator phases: a Mott insulator phase and $\rho$ CDW phases. For non-integer densities larger than one, several CDW phases appear~\cite{Cruz-EPJB16}.\par 
\textcolor{red}{Superfluid-to-Mott insulator transitions with spinor bosons confined in optical lattices have been observed~\cite{SoltanPanahi-NP11,Stamper-RMP13,Jiang-PRA16}, however, neither the possibility of driving this transition through a structural deformation of the unit cell nor the topological character of the different phases have been considered in the experiments, and the question about the consequences of considering the internal degrees of freedom on the critical points of inhomogeneous lattices arises. In order to stimulate experiments and give a first idea about what to expect, we address this issue in this manuscript. In a pioneer study,} Wagner {\it et al.} used the mean-field approximation to study spinless and spin-1 bosons in an intercalated potential and concluded that spin-dependent interactions change the occupation numbers of individual lattice sites~\cite{Wagner-PRA12}. Motivated by the above scenario, we went beyond the mean-field approximation and used the density matrix renormalization group method~\cite{White-PRL92,Hallberg-AP06} to study spin-1 bosons in a superlattice potential\textcolor{red}{, considering an effective  antiferromagnetic local interaction}. We found that the spin-dependent interaction favors insulator phases for integer densities, which can arise from a finite or zero value of the spin-dependent strength, depending on the particular value of the energy shift. The CDW phases for semi-integer densities decrease and tend to disappear with the spin-dependent strength. Phase diagrams as a function of the energy shift or the spin-dependent interaction were calculated at the thermodynamic limit.\par
\textcolor{red}{The above results can stimulate the quantum simulation of higher spin chains in cold atoms setups and the possible connection between internal degrees of freedom and the topological character of the diverse phases~\cite{Li-PRB15}.}\par 
The outline of this manuscript is as follows: In Sec. \ref{model} we explain the Hamiltonian model that describes spin-1 bosons in a superlattice potential, and the limit case without a kinetic energy term (atomic limit) is analyzed, leading to the first phase diagrams. The chemical potential at the thermodynamic limit is calculated numerically and shown in Sec \ref{results}, where the main phase diagrams appear. Finally, in Sec. \ref{conclusions} we summarize our principal results and state our conclusions.\par  
\begin{figure}[t]
\centering
\includegraphics[width=0.45\textwidth]{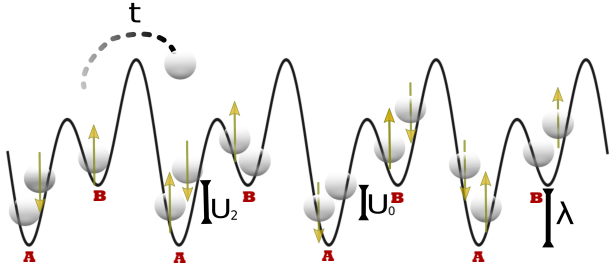}
\caption{Schematic of the setup related to Hamiltonian (\ref{eq1}). Spin-1 bosons in a superlattice optical potential (AB chain). $t$ is the hopping parameter to the nearest neighbor,  the on-site repulsion interaction is given by $U_{0}$, $U_{2}$ is the exchange interaction, and $\lambda$ is the energy shift between sites. The arrows indicate the hyperfine state of each atom $F^z=1$(up), $F^z=-1$ (down), and $F^z=0$ (none).}
\label{fig1}
\end{figure} 

\section{\label{model} Model}
A system of spinor bosons with spin $F=1$ in one dimension can be described by considering a kinetic term with a hopping parameter between the neighbor sites $t$, a local repulsion interaction of strength $U_{0}$, an effective local interaction due to the spin with strength $U_{2}$ and finally the local potential undergone by each boson in the lattice. The Hamiltonian associated with the above system is given by 
\begin{eqnarray}
\hat{H}= & &-t\sum_{<i,j>,\sigma}\left(\hat{a}_{i,\sigma}^{\dagger}\hat{a}_{j\sigma}+\hat{a}_{j,\sigma}^{\dagger}\hat{a}_{i\sigma}\right)+\frac{U_{0}}{2}\sum_{i}\hat{n}_{i}(\hat{n}_{i}-1)\nonumber\\
& &+\frac{U_{2}}{2}\sum_{i}(\hat{F}_{i}^{2}-2\hat{n}_{i})+\sum_{i}\lambda_{i}\hat{n}_{i}-\mu\sum_{i}\hat{n}_{i},
\label{eq1}
\end{eqnarray}
\noindent $\hat{a}_{i,\sigma}^{\dagger}(\hat{a}_{i,\sigma})$ being the creation (annihilation) operator of a boson at site $i$ in the magnetic sublevels $\sigma=1,~0,~-1$. $\hat{n}_{i}$ is the number operator and $\hat{F}_{i}=\sum_{\sigma,\sigma '}\hat{a}^{\dagger}_{i,\sigma}\mathbf{T}_{\sigma,\sigma '} \hat{a}_{i,\sigma '}$ is the spin operator with $\mathbf{T}_{\sigma,\sigma '}$ being the $S$-1 Pauli matrices. $\mu$ represents the chemical potential, and $\lambda_i$ quantifies the local external potential undergone by the bosons, which is periodic with an unitary cell $AB$, such that $\lambda_{i}=0$ if the site is A and $\lambda_i=\lambda$ if the site is B ( see Fig. \ref{fig1}). Note that 
\textcolor{red}{$-1<\dfrac{U_{2}}{U_{0}}=\dfrac{a_{2}-a_{0}}{a_{0}+2a_{2}}<0.5$,} where $a_{n}$ are the scattering lengths. \textcolor{red}{For an antiferromagnetic ($U_{2}>0$) homogeneous chain, an even-odd asymmetry between the Mott lobes was found, while in the ferromagnetic case ($U_{2}<0$), both superfluid and insulator phases exhibit magnetic quasi-long-range order. Although the ferromagnetic case may be interesting, we chose an antiferromagnetic spin-dependent interaction in our inhomogeneous chain, and all our results will be limited to this case.}\par
To explore the ground state of spinor bosons in an AB chain, we first consider the atomic limit ($t=0$ in the Eq. (1)) in which the energy for 
$n_{i}$ particles in the unit cell is given by:
\begin{eqnarray}
E_{0}(n_{i})= \frac{U_{0}}{2}n_{i}(n_{i}-1)+\frac{U_{2}}{2}\left(\braket{F_{i}}^{2}-2n_{i}\right)\nonumber\\+\lambda_{i}n_{i}-\mu n_{i},
\end{eqnarray}
\noindent where $\braket{\hat{F}^{2}}=0$ for an even number of particles and $\braket{\hat{F}^{2}}=2$ if the number of particles is odd, when an antiferromagnetic interaction is considered.\par
\begin{figure}[t] 
\includegraphics[width=18pc]{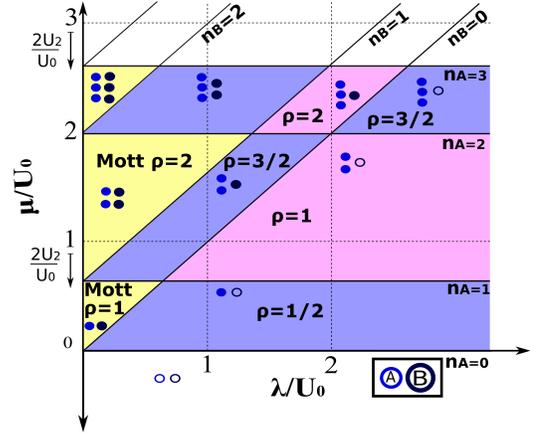} 
\caption{\label{fig2} Phase diagram at the atomic limit of spin-1 bosons confined in an AB chain. The distribution of the particles in sites A and B into the insulator regions is illustrated. Filled circles indicate that there is a particle, and empty circles that there are no particles. The left (right) group of circles belongs to the site A (B). It can be seen that only lines with odd $n_{i}$ change by a downward shift in the vertical axis for $\frac{2U_{2}}{U_{0}}$, and the size regions change, keeping the critical point for $\rho=3/2$} 
\end{figure} 
At the atomic limit, the ground state is characterized by a particular occupation of each site of the unit cell, and a change of state can happen when the parameters vary. This change of state takes place when $E_{0}(n_{i}+1)-E_{0}(n_{i})$ is equal to zero; hence the boundaries between different states are given by lines of the chemical potential ($\frac{\mu}{U_{0}}$) in terms of $\frac{\lambda_{i}}{U_{0}}$ and $\frac{U_{2}}{U_{0}}$ in the following way:
\begin{equation}
\label{atl1}
\frac{\mu}{U_{0}}= \frac{\lambda_{i}}{U_{0}}-\frac{2U_{2}}{U_{0}}+n_{i},
\end{equation}
\noindent when $n_{i}$ is odd, and
\begin{equation}
\label{atl2}
\frac{\mu}{U_{0}}= \frac{\lambda_{i}}{U_{0}}+n_{i},
\end{equation}
\noindent if $n_{i}$ is even.\par 
\begin{figure}[t] 
\includegraphics[width=18pc]{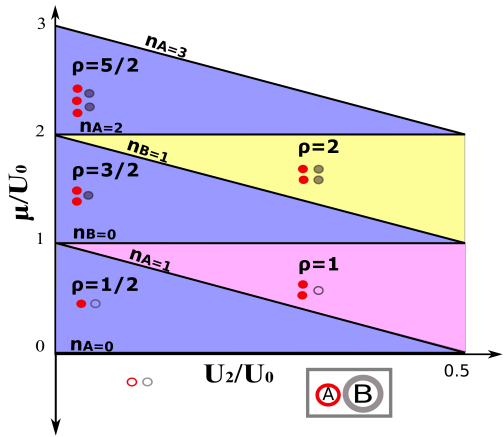} 
\caption{\label{fig3}$t=0$ phase diagram in the plane chemical potential versus spin-dependent interaction. The symbols are similar to those of Fig.~\ref{fig2}. Here, we fix $\lambda/U_{0}=1$. } 
\end{figure} 
The ground state phase diagram in the plane chemical potential versus the energy difference is shown in Fig.~\ref{fig2}. Here, we consider that $2U_2/U_0=0.38$, and the horizontal (inclined) lines set the border for which the number of particles at site A(B) changes according to the relations (\ref{atl1}) and (\ref{atl2}). The number of particles in a unit cell is shown, such that left (right) points correspond to the occupation of the site A (B). This figure shows us that the ground state can be a Mott insulator or a charge density wave (CDW) state and that transitions between them can be driven by the energy shift ($\lambda$) or the spin-dependent interaction ($U_2$). For instance, phase transitions from Mott to CDW insulator are obtained for integer global densities, while for semi-integer densities, only transitions between different CDW insulators are possible. The critical point for the latter transitions does not depend on the spin-dependent interaction $U_2$, whereas for the former the critical point will depend on the density and the spin-dependent interaction. We observe that for $\rho=1$, the critical point is located at $\lambda=U_{0}-2U_{2}$, while for $\rho=2$ it is at $\lambda=U_{0}+2U_{2}$.\par 
The special case when $\lambda/U_{0}=1.0$ at the atomic limit is depicted in Fig. \ref{fig3}. In the vertical axis are the values of the chemical potential $\mu$, while in the horizontal axis the values of $U_{2}$ are displayed, both in terms of $U_{0}$. Note that according to equations (\ref{atl1}) and (\ref{atl2}), the phase diagram is different. This does not suggest transitions from Mott to CDW for integer densities. The Mott insulator for $\rho=1$ does not appear, and semi-integer CDW regions are separated by insulator Mott or CDW regions with integer densities. Also, this phase diagram suggests that the insulator regions for integer densities appear starting at $U_{2}=0$.\par 
For larger values of $\lambda_{i}/U_{0}$, interesting things happen at the atomic limit. For instance, no Mott insulator phases appears for  
$\lambda_{i}/U_{0}=2$, and for lower values a CDW to Mott transition with a density $\rho=2$ is expected for a finite value of spin-dependent interaction $U_{2}$ (not shown).\par
\begin{figure}[t]
\includegraphics[width=8cm]{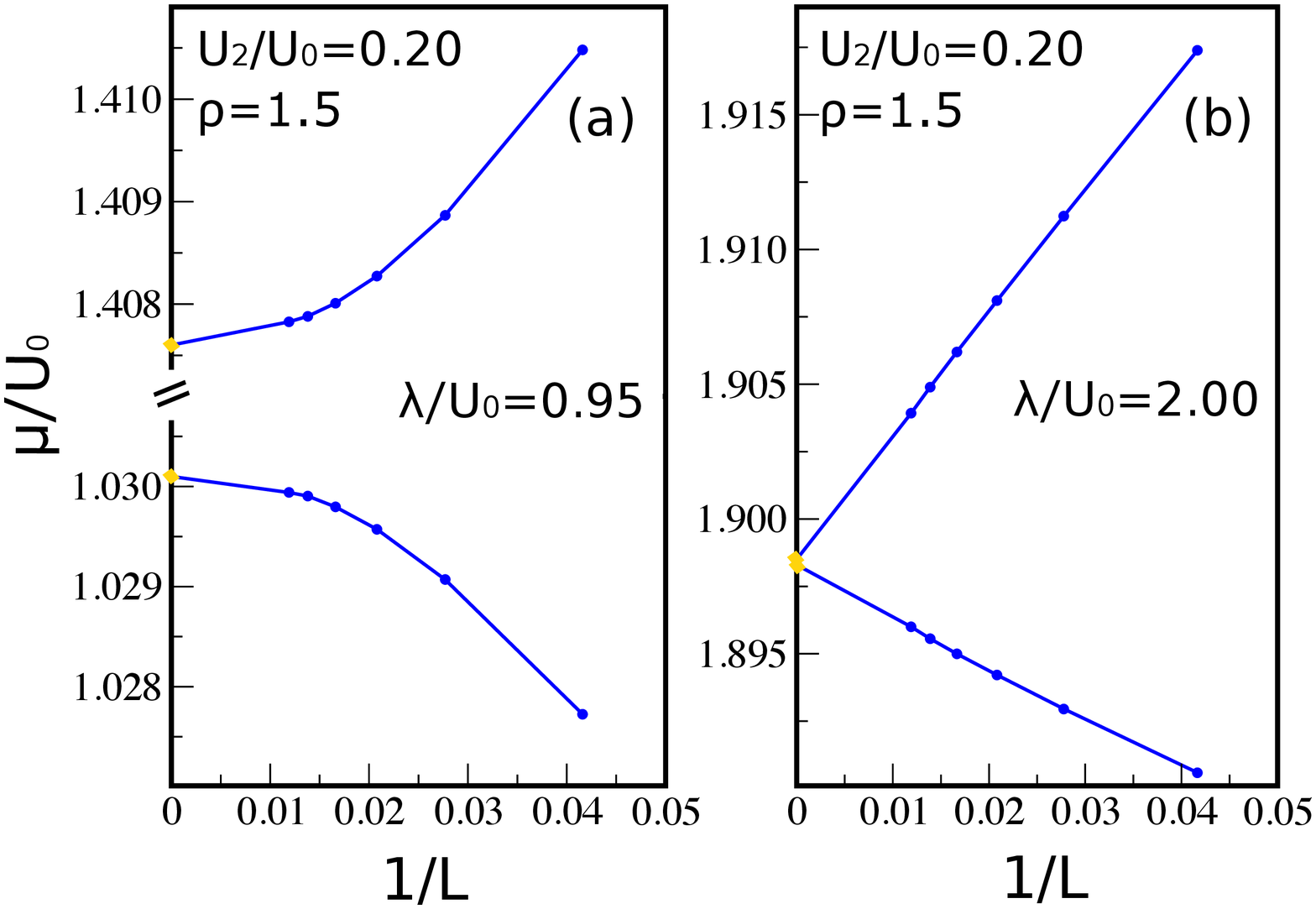}
\caption{\label{fig4} Chemical potential versus the inverse of the lattice size for a system of spinor bosons with $\rho=3/2$ and $U_{2}/U_{0}=0.2$. Two different values of energy shift were considered $\lambda/U_{0}=0.95$ (a) and $\lambda/U_{0}=2.0$ (b). At the thermodymanic limit, a finite charge gap ($\Delta\neq 0$) is obtained in (a), while it vanishes in (b).}
\end{figure}
\section{\label{results} Numerical  Results} 
The atomic limit of the Hamiltonian (\ref{eq1}) shows us that as spinor bosons are confined in the AB chain, diverse insulator phases can appear and transitions between them can occur. However, beyond the atomic limit, the quantum fluctuations will modify the above picture, and a numerical analysis becomes important. We chose the density matrix renormalization group (DMRG) method to study the Hamiltonian (\ref{eq1}). The calculations were carried out by keeping $m=350$ states, where the accuracy of the discarded weight was $10^{-5}$ in the worst case. \textcolor{red}{The ground state energy difference between successive sweeps (from left to right) was on the order of 0.01. The hopping and the local interaction parameters were fixed, and we considered the values $t=1$ and $U_{0}/t=10$, respectively.}\par
%
\begin{figure}[t]
\includegraphics[width=8cm]{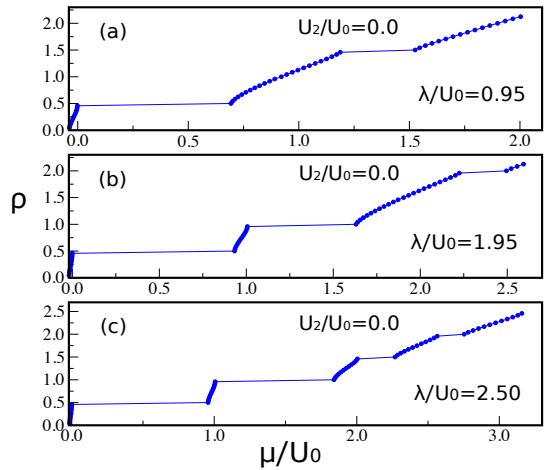}
\caption{\label{fig5} Density $\rho=N/L$ versus the chemical potential for spinor bosons without spin-dependent interaction ($U_{2}=0$). Here, two different values of the energy shift were considered, $\lambda/U_{0}= 0.95$ and $1.95$. The chemical potential values correspond to those at the thermodynamic limit.}
\end{figure}
It is well-documented in the literature that the insulator phases are characterized by a finite charge gap $\Delta=\mu_{p}-\mu_{h}$ at the thermodynamic limit, $\mu_{p}$ ($\mu_{h}$) being the chemical potential for adding (removing) a particle, which are given by:
\begin{equation}
\mu_{p}(L)=E(N+1,L,F^{z})-E(N,L,F^{z}),
\label{mup}
\end{equation}
 and
 \begin{equation}
\mu_{h}(L)=E(N,L,F^{z})-E(N-1,L,F^{z}),
\label{muh}
\end{equation}
\noindent where $E(N,L,F^{z})$ is the system energy with $N$ particles, $L$ sites, and spin proyection $F^{z}$.\par
In Fig.~\ref{fig4}, we show the evolution of the  chemical potential for adding  and removing a particle as a function of the inverse of the lattice size, for a system of spinor bosons with a global density $\rho=3/2$ and \textcolor{red}{an antiferromagnetic} spin-dependent interaction $U_{2}/U_{0}=0.2$. We observed that the $\mu_{p}$ ($\mu_{h}$)) decreases (increases) monotonously as the system size grows; however, for $\lambda/U_{0}=0.95$ (Fig. \ref {fig4}(a)), both quantities tend to different values at the thermodynamic limit ($1/L \rightarrow 0$), indicating that for these conditions the system has a finite charge gap $\Delta/U_{0}=0.38$ and the ground state is a CDW as predicted at the atomic limit (Fig. \ref {fig2}).\par 
On the other hand, Fig. \ref {fig4}(b) shows the case for $\lambda/U_{0}=2.0$. The tendency of both chemical potentials to a unique value at the thermodynamic limit can clearly be seen, namely $\mu/U_{0}=1.90$ and $\Delta/U_{0}=0$. Therefore, there is a superfluid phase separating the CDW phases with global density $\rho=3/2$ as the energy shift increases, keeping the others parameters constant. Clearly, outside of the atomic limit we obtain quantum phase transitions between superfluid and insulator phases, which are absent in Figs.~\ref {fig2} and  ~\ref {fig3}. The above fact leads us to study the phase diagram of spinor bosons confined in an AB chain far away of the atomic limit.\par
\begin{figure}[t]
\includegraphics[width=8cm]{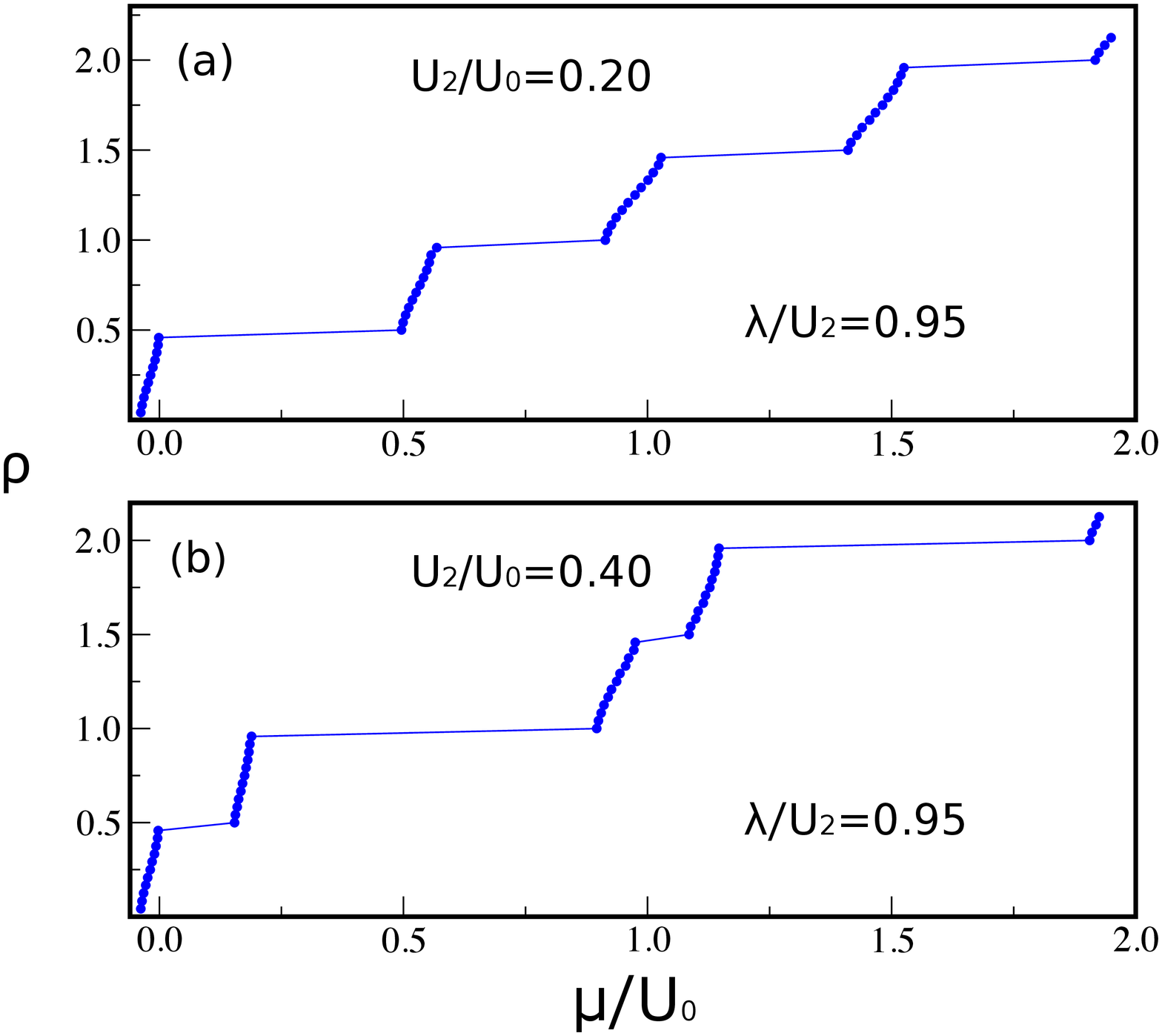}
\caption{\label{fig6} Density $\rho=N/L$ versus the chemical potential for spinor bosons. The energy shift was set at $\lambda/U_{0}= 0.95$ and different non-zero values of the spin-dependent interaction were considered. The chemical potential values correspond to those at the thermodynamic limit.}
\end{figure}
So far, our results suggest that the ground state can be gapped or gapless; hence we must explore the evolution of the chemical potential at the thermodynamic limit as the number of spinor bosons increases for different sets of parameters. For spinor bosons without spin-dependent interaction $(U_{2}=0)$, a situation that corresponds to the spinless bosons case, we show in Fig.~\ref{fig5} the density $\rho=N/L$ as a function of the chemical potential. It is well-known that for a homogeneous lattice $\lambda/U_{0}=0$, the ground state is superfluid for non-integer densities, displaying a continuous increase in the density vs the chemical potential curve. But for integer densities the ground state is a Mott insulator, and horizontal discontinuity (plateaus) appear in the curve, where the width of the plateau informs the value of the gap. For bosons in an AB chain with $\lambda/U_{0}=0.95$ (see Fig.~\ref{fig5} (a)), the Mott insulator plateaus for integer densities disappear, and instead plateaus are observed for semi-integer densities, where the ground state is a CDW with the particular unit cell filling $\{A=1,B=0\}$ for $\rho=1/2$ and  $\{2,1\}$ for $\rho=3/2$, according to the phase diagram Fig.~\ref {fig2}. \textcolor{red}{Note that negative values of the chemical potential appear. This means that the ground-state energy decreases in order to increase the number of particles, keeping the entropy constant~\cite{Cook-AJP95}}. For larger values of the energy shift ($\lambda/U_{0}=1.95$), the plateau for $\rho=3/2$ disappears, the CDW phase for $\rho=1/2$ is maintained and enlarged and the insulator phases for integer densities reappear(see Fig.~\ref{fig5} (b)). However, it is expected that the latter insulator phases will not correspond to Mott ones; instead they will be CDW insulator phases (Fig.~\ref {fig2}). Note that plateaus for all semi-integer and integer values of the density will appear as the energy shift grows (see Fig.~\ref{fig5} (c)). In conclusion, the superlattice potential leads to the emergence of diverse CDW insulator phases for semi-integer and integer densities and drives quantum phase transitions between insulator phases (Mott or CDW) and superfluid ones for particular values of the parameters. The above results and phase diagrams have previously been obtained by some authors, using diverse numerical methods~\cite{Dhar-PRA11a,Dhar-PRA11b,Singh-PRA12,Cruz-EPJB16}.\par
\begin{figure}
\includegraphics[width=8.5cm]{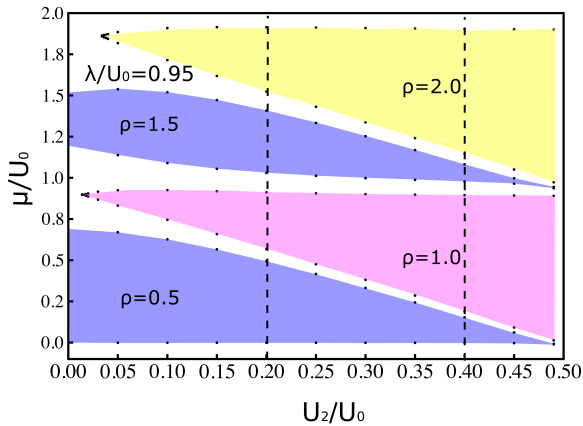}
\caption{\label{fig7} Phase diagram for spinor bosons in a superlattice type $AB$ at the thermodynamic limit. Here, the energy shift was set at 
$\lambda/U_{0}=0.95$. The white regions correspond to superfluid phases, the blue regions indicate CDW phases for semi-integer densities, and  the Mott insulator phase is represented in yellow, whereas the CDW phase for integer densities is in pink. The dashed lines correspond to the  chemical potential as the number of boson increases, shown in Fig.~\ref {fig6}. The points represent the boundaries of the insulator phases calculated with DMRG.}
\end{figure}
The effective description of spinor bosons provides a local spin-dependent interaction term, whose effect on the ground state of spinor bosons in an AB chain we want to study. In Fig.~\ref{fig6} we display the density $\rho=N/L$ versus the chemical potential for bosons in an AB chain with $\lambda/U_{0}= 0.95$ and the spin-dependent interactions $U_{2}/U_{0}=0.2$ (a) and $U_{2}/U_{0}=0.4$ (b). For the energy shift $\lambda/U_{0}= 0.95$ and without spin-dependent interaction, we show in Fig.~\ref{fig5} (a) that only for semi-integer densities are there  insulator phases, namely CDW ones. When we turn on the spin-dependent interaction (see Fig.~\ref{fig6} (a)), we observ that the plateaus for integer densities reappear, which indicates that a quantum phase transition from a superfluid to an insulator phase is driven by the spin-dependent interaction for a given AB chain. Remember that the atomic limit says that the critical point is $U_{2}/U_{0}=0$ for both integer densities. According to the phase diagram Fig.~\ref {fig3}, we note that the plateaus for semi-integer densities are present, but their width has changed, in particular the plateau for $\rho=3/2$. As the spin-dependent interaction increases, the plateaus for integer densities are larger, while the ones for semi-integer densities decrease quickly, which follows the atomic limit results, and therefore we expected that the insulator phases for semi-integer densities would tend to disappear as $U_{2}/U_{0}\rightarrow 0.5$ (Fig.~\ref{fig6} (b)).\par
\begin{figure}
\includegraphics[width=8.5cm]{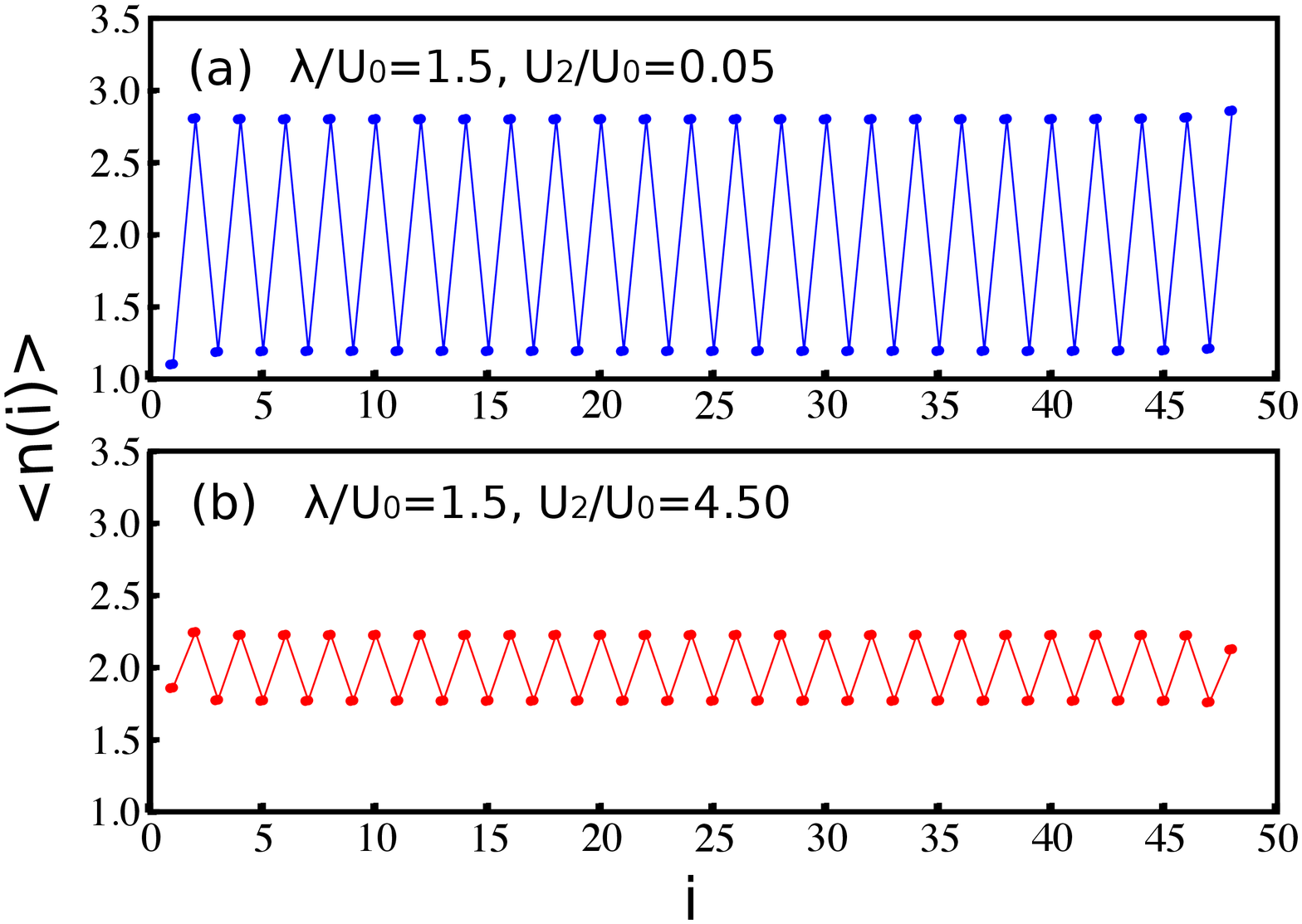}
\caption{\label{fig8} On-site number density plotted against lattice site index at density $\rho=2$ and values smaller or larger than the critical point. Here we consider an AB chain with $\lambda/U_{0}= 1.5$. The top panel corresponds to $U_{2}/U_{0}=0.05$ and the bottom one to  $U_{2}/U_{0}=0.45$. Open boundary conditions were used. The lines are visual guides.}
\end{figure}
In Fig.~\ref{fig6}, we see that insulator phases for integer densities are favored by the spin-dependent interaction, but what kind of insulator are they? We calculated the density profile and found that for $\rho=1$, the particle distribution in the unit cell is $\{2,0\}$, hence the ground state is a CDW for this density. However, for $\rho=2$ the average occupation per site is $<n(i)>\approx 2.0$, which indicates that the ground state is a Mott insulator (not shown).\par 
The phase diagram of spinor bosons confined in an AB chain with an energy shift $\lambda/U_{0}= 0.95$ is shown in Fig.~\ref{fig7}. The points correspond to the thermodynamic limit of the chemical potential for different values of spin-interaction strength ($U_{2}/U_{0}$) and give the boundaries of the insulator regions. Comparing it with the phase diagram at the atomic limit (Fig.~\ref {fig3}), we note that the kinetic energy generates superfluid regions that separate the insulator ones, prohibiting a quantum phase transition between them even without spin-dependent interaction. The predicted insulator regions at the atomic limit are present in the phase diagram. For small values of $U_{2}/U_{0}$, the predominant regions are CDW for semi-integer densities $\rho=1/2$ and $\rho=3/2$ with charge distribution in the unit cell $\{1,0\}$ and $\{2,1\}$, respectively, although the charge gap for the insulator with  density $\rho=3/2$ is smaller than that predicted. As the spin-interaction strength grows the insulator regions for semi-integer densities decrease and tend to disappear as $U_{2}/U_{0}\rightarrow 0.5$, as can be seen in the figure, where the extreme value $U_{2}/U_{0}=0.49$ was considered. We found that the insulator regions for integer densities appear for different non-zero values of the spin-dependent interaction and quantum phase transitions from a superfluid to an insulator phase take place at the critical values $U_{2c}/U_{0}=0.012$ and $U_{2c}/U_{0}=0.032$ for the densities $\rho=1$ and $\rho=2$, respectively. As mentioned before, the insulator region for the density $\rho=1$ corresponds to a CDW phase with a distribution of particles in the unit cell $\{2,0\}$, whereas there is a Mott insulator phase for $\rho=2$. The insulator phases for integer densities grow as the spin-interaction strength increases, dominating the phase diagram at the limit $U_{2}/U_{0}\rightarrow 0.5$. For both integer densities, we observed that the upper border quickly assumes an almost constant value, displaying a horizontal line, while the lower border evolves almost linearly , which reminds us of the prediction of the atomic limit (see Fig.~\ref {fig3}).\par
\begin{figure}
\includegraphics[width=9cm]{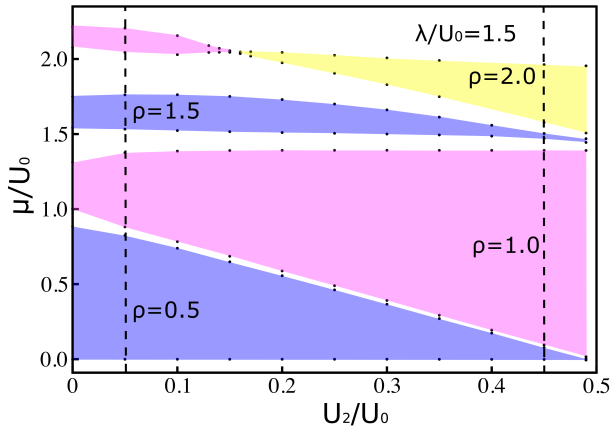}
\caption{\label{fig9} Phase diagram for spinor bosons in a superlattice type $AB$ at the thermodynamic limit. Here, the energy shift was set at 
$\lambda/U_{0}=1.5$. The white regions correspond to superfluid phases, and the blue regions indicate CDW phases for semi-integer densities. The Mott insulator phase is represented in yellow, whereas the CDW phase for integer densities is in pink. The dashed lines correspond to the density profiles shown in Fig.~\ref {fig8}. The points represent the boundaries of the insulator phases calculated with DMRG.}
\end{figure}
Previous studies of spinless bosons confined in an AB chain show that quantum phase transitions take place at an energy shift around multiples of the local repulsion $U$. For this reason, we chose $\lambda/U_{0}= 0.95$ to do our previous analysis. But what happens for larger values of the energy shift? In Fig.~\ref{fig8}, the on-site number density profile is shown for $\lambda/U_{0}= 1.5$ and a global density $\rho=2$.  Far  from our previous result, we obtained a pattern of three particles in site A and one in site B when the spin-dependent interaction is $U_{2}/U_{0}=0.05$. Hence for this set of parameters, the insulator regions for integer densities correspond to CDW phases. The on-site number density profile for $U_{2}/U_{0}=0.45$ oscillates around the double occupancy, which indicates that the ground state is a Mott insulator. The above results suggest that for finite values of the spin-interaction strength, a quantum phase transition takes place for a fixed density $\rho=2$ and $\lambda/U_{0}= 1.5$.\par
\begin{figure}
\includegraphics[width=9cm]{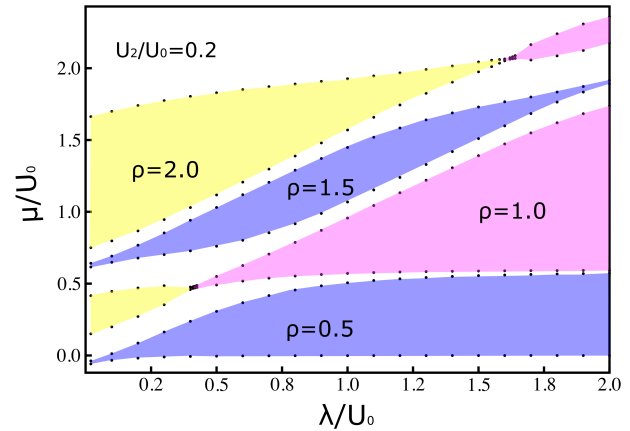}
\caption{\label{fig10} Phase diagram in the plane of chemical potential versus the energy shift for spinor bosons with spin-dependent interaction $U_{2}/U_{0}=0.2$. The chemical potential values are calculated at the thermodynamic limit. The white regions correspond to superfluid phases, and the blue regions indicate CDW phases for semi-integer densities. The Mott insulator phase is represented in yellow, whereas the CDW phase for integer densities is in pink. The points represent the boundaries of the insulator phases calculated with DMRG.}
\end{figure}
In Fig.~\ref{fig9}, the chemical potential values at thermodynamic limit as a function of the spin-dependent interaction are shown. In this phase diagram, superfluid, Mott insulator, and CDW insulator phases appear, where again superfluid phases surround and separate the insulators phases. In the current case, we consider $\lambda/U_{0}= 1.5$ and observe that the \textcolor{red}{CDW} insulator phases for semi-integer densities preserve the main characteristics discussed in Fig.~\ref{fig7} for an energy shift $\lambda/U_{0}= 0.95$. A new fact shown in the current phase diagram is that without spin-dependent interaction, there is a finite charge gap for integer densities, which indicates that there will be no quantum transitions near $U_{2}/U_{0}=0$, and both insulator regions are CDW, in accordance with the spinless case. Note that for density $\rho=1$, 
the overall evolution of the boundaries is similar to those discussed in Fig.~\ref{fig7}, except for the fact that this phase exists without spin-dependent interaction, and this particular charge distribution in the unit cell $\{2,0\}$ is determined by the antiferromagnetic interaction that dominates the phase diagram. However, for a global density  $\rho=2$, Fig.~\ref{fig8} suggests a quantum transition for a finite value of $U_{2}/U_{0}$, and we observe two regions for this density. The first region corresponds to a CDW insulator, and a quantum phase transition from CDW insulator to superfluid takes place. A second quantum phase transition from superfluid to Mott insulator occurs, and the latter phase survives up to the limit $U_{2}/U_{0}\rightarrow 0.5$. These two quantum phase transitions are very close: we estimate that they are in the range $U_{2}/U_{0}=(0.150,0.155)$.\par 
For larger values of the energy shift (for instance $\lambda/U_{0}= 2.0$), we expected a phase diagram with insulator phases for integer densities and only one CDW insulator phase for semi-integer ones, i.e. there is no CDW insulator with density $\rho=3/2$ for any value of the 
spin-interaction strength, which is accordance with Fig.~\ref{fig4} (b). It is also predicted that the area of the CDW insulator with density $\rho=2$ will increase and that the critical points will move to larger values of the spin-dependent interaction.\par
Without spin-dependent interaction, the well-known phase diagram in the plane chemical potential versus the energy shift shows Mott-superfluid-CDW transitions as the energy shift increases for integer densities, which take place around $\lambda/U_{0}=1$. Also, CDW-superfluid-CDW  transitions occur for the density $\rho=3/2$ around $\lambda/U_{0}=2$. The unit cell configuration for the first CDW is $\{2,1\}$, and for the second one it is  $\{3,0\}$. Finally, a growing and ever present CDW insulator phase for  $\rho=1/2$ was reported~\cite{Dhar-PRA11a,Dhar-PRA11b,Singh-PRA12,Cruz-JPCS16,Cruz-EPJB16}. Turning on the spin-dependent interaction ($U_{2}/U_{0}=0.2$), we obtain a phase diagram of the chemical potential as a function of the energy shift with the same phases as the phase diagram with $U_{2}/U_{0}=0$ (see Fig.~\ref {fig10}). We observe that the overall behavior of the CDW insulator phases for semi-integer densities is the same as that reported previously, a quantum phase transition from CDW insulator to superfluid taking place around $\lambda/U_{0}=2$ for a density $\rho=3/2$, in accordance with Figs.~\ref{fig2} and ~\ref{fig4}. The great changes occur for the integer densities; here the Mott-superfluid-CDW transitions do not occur around $\lambda/U_{0}=1$. For $\rho=1$, the Mott insulator region decreases, and the transitions take place around $\lambda/U_{0}\sim 0.48$, whereas the spin-dependent interaction favors the Mott insulator phase for $\rho=2$, moving the critical region to $\lambda/U_{0}\sim 1.61$. As expected the above critical points are displaced with respect to the atomic limit forecast, due to the quantum fluctuations.\par 
%


%
\section{\label{conclusions} Conclusions}
We investigated the role of the spin degree of freedom on the ground state of spinor bosons confined in a superlattice potential with a unit cell 
$\{A,B\}$ that has an energy shift $\lambda$ between its sites. Taking into account the local repulsion ($U_{0}$) and the effective spin interaction term ($U_{2}$) for spin-1 bosons, we obtained that the spatial structure of the lattice generates a charge redistribution that leads to charge density wave phases.\par 
Studying the Hamiltonian model at the atomic limit (without a kinetic term) and calculating the chemical potential with the density matrix renormalization group method, we built phase diagrams in the plane chemical potential as a function of the energy shift or the spin-dependent  interaction for several sets of parameters. Fixing the energy shift, we found that the CDW phases for semi-integer densities decrease and tend to disappear as $U_{2}/U_{0}\rightarrow 0.5$. Also, the spin-dependent interaction favors insulator phases for integer densities, which can arise from a finite or zero value of the spin-dependent strength depending on the particular value of the energy shift. For a global density $\rho=2$ and $\lambda>U_{0}$ CDW, insulator-superfluid-Mott insulator quantum phase transitions can be driven for finite values of the spin-dependent interaction.\par 
The phase diagram as a function of the energy shift shows that the CDW insulators with semi-integer densities preserve their main characteristics, as in the spinless case ($U_{2}=0$). However, the Mott insulator-superfluid-CDW insulator quantum phase transitions do not occur around $\lambda\sim U_{0}$; in particular, the critical point will be at $\lambda < U_{0}$ for a global density $\rho=1$ and at $\lambda > U_{0}$ for $\rho=2$.\par 
A more precise determination of the critical points was not possible due to the large local Hilbert space (considering up to three bosons per site, the local dimension is twenty) and our limited computational resources, but we believe that the overall physics reported in this paper will not change.\par 
\textcolor{red}{A possible experimental implementation of the model studied here may consider $^{23}Na$ atoms, for which $U_{2}\simeq0.04U_0$ and the superfluid-to-Mott-insulator transitions have been observed in cold atoms setups~\cite{Jiang-PRA16}. Also note that diverse lattice configurations and inhomogeneities have been achieved using optical lattices~\cite{Roati-N08,Atala-NP13,Windpassinger-RPP13,DiLiberto-NC14,Messer-PRL15}, which allows predicting that the model considered could be implemented and the predicted transitions would be observed.}\par 
\textcolor{red}{One perspective of this contribution is to consider the ferromagnetic case, which corresponds to confining Rb atoms in an inhomoheneous optical lattice, and we will try to connect our results with a very recent experimental measures of a superlattice with a three-site unit cell~\cite{Anderson-ArX19}.}\par


\section*{Acknowledgments}
J. S.-V. is thankful for the support of DIEB- Universidad Nacional de Colombia (Grant No. 41402). G.J. Cruz is thankful for the support of the Marshall Plan Scholarship Program (Grant No. 858 1092 24 20 2018) and the hospitality of the Johannes Kepler Universit\"{a}t Linz, Physics Department, where part of this work was done.

\bibliography{Bibliografia}

\end{document}